\documentclass[aps,preprint,superscriptaddress,amsmath,amssymb,showpacs]{revtex4-1}
\usepackage[utf8]{inputenc}
\usepackage{amsmath}
\usepackage{siunitx}
\usepackage{graphicx}
\usepackage{sidecap}
\usepackage[dvipsnames]{xcolor}
\graphicspath{{.}{./figures-main/}}
\usepackage{nameref,zref-xr}
\zxrsetup{toltxlabel}
\usepackage[colorlinks,linkcolor=blue,citecolor=blue,anchorcolor=blue]{hyperref}
\usepackage{cleveref}
\usepackage{textcomp}
\definecolor{dark-green}{rgb}{0.000, 0.500, 0.000}
\usepackage{lineno}
\begin{document}

\title{Persistent half-metallic ferromagnetism in a (111)-oriented manganite superlattice}

\author{Fabrizio Cossu}
\affiliation{Department of Physics and Institute of Quantum Convergence and Technology, Kangwon National University -- Chuncheon, 24341, Korea}
\affiliation{Asia Pacific Center for Theoretical Physics -- Pohang, 37673, Korea}

\author{Heung-Sik Kim}
\email[H.-S. Kim: ]{heungsikim@kangwon.ac.kr}
\affiliation{Department of Physics and Institute of Quantum Convergence and Technology, Kangwon National University -- Chuncheon, 24341, Korea}

\author{Biplab Sanyal}
\affiliation{Department of Physics and Astronomy, Uppsala University, Box 516, SE-75120, Uppsala, Sweden}

\author{Igor Di Marco}
\email[I.~Di Marco: ]{igor.dimarco@apctp.org}
\affiliation{Asia Pacific Center for Theoretical Physics -- Pohang, 37673, Korea}
\affiliation{Department of Physics and Astronomy, Uppsala University, Box 516, SE-75120, Uppsala, Sweden}
\affiliation{Department of Physics, POSTECH, Pohang, 37673, Korea}

\date{\today}

\begin{abstract}
	We employ electronic structure calculations to show that a (111)-oriented (LaMnO$_3$)$_{12}\vert$(SrMnO$_3$)$_{6}$ superlattice retains a half-metallic
	ferromagnetic character despite its large thickness. We link this behaviour to the strain and the octahedral connectivity between the layers. This also
	gives rise to breathing modes, which are coupled to charge and spin oscillations, whose components have a pure $e_{g}$ character. Most interestingly,
	the magnetization reaches its maximum value inside the LaMnO$_3$ region and not at the interface, which is fundamentally different from what observed
	for the (001) orientation. The inter-atomic exchange coupling shows that the magnetic order arises from the double-exchange mechanism, despite competing
	interactions inside the SrMnO$_3$ region. Finally, the van Vleck distortions and the spin oscillations are crucially affected by the variation of Hund's
	exchange and charge doping, which allows us to speculate that our system behaves as a Hund's metal, creating an interesting connection between manganites
	and nickelates.
\end{abstract}


\maketitle 

\section{Introduction}
  Due to continuous advances in molecular beam epitaxy \cite{prakash-wileybook.ch26,ismailbeigi-NatRevMat2017} and pulsed laser deposition
 \cite{christen-JPCM2008,aruta-APL2010,eres-PRL.117.206102,ismailbeigi-NatRevMat2017,yao-apA2019,lorenz-wileybook.eap810,koster-JSNM2020},
 perovskites (and the more general Ruddlesden-Popper family, A$_{n+1}$B$_{n}$O$_{3n+1}$, where $n = \infty$ for perovskites) have been objects
 of undying attention in the scientific community \cite{gao_x-AdvSci2019,chen_h-JPCM2017,righetto-APR2021}. In the last years, this trend has
 been further accelerated by advances in defect engineering, which has improved the perspective of practical applications
 \cite{koster-JSNM2020,li_w-MH2020}. Despite their apparent simplicity, bulk materials exhibit a variety of ground states, driven by the strong
 interplay of different degrees of freedom \cite{dagotto-Science2005,ngai-AnnuRevMR2014,tokura-Science2000,cheong_sw-nmat2007,schlom-MRSBull2014}.
 The symmetry breaking at interfaces \cite{flint-PRM.3.064401,grutter-NanoL2016,flint-JAP2014,keshavarz-PRB.95.115120} leads to even more exciting
 phenomena, such as flat bands \cite{ruegg-PRB.85.245131}, anisotropic conductivity \cite{annadi-ncomm2013}, magnetic anisotropy
 \cite{kim_b-PRB.89.195411,asaba-PRB.98.121105}, exchange bias \cite{gibert-nmat2012}, spin-glass \cite{huangfu_s-PRB.102.054423}, electronic
 quantum confinement \cite{yoo-PRB.93.035141,liu_j-PRB.83.161102,dong-PRB.87.195116}, unconventional superconductivity \cite{li_d-Nature2019},
 topologically protected edge states \cite{doennig-PRL.111.126804,ruegg-PRB.88.115146}, unexpected metallicity
 \cite{li_c-APL2013,myself-PRB.87.214420,myself-AMI2014,zubko-AnnuRevCMP2011,hwang-nmat2012} and tunable quantum phase transitions
 \cite{cdam-adma2021,hwang-nmat2012,pauli_sa-JPCM2008}.

  Much research on interfaces and surfaces has been focused on mixed-valence manganites. Bulk materials are primarily known for their colossal magnetoresistance
 \cite{imada-RMP.70.1039}, which is favoured by a disordered solid mixture of Mn$^{+3}$ and Mn$^{+4}$ ions \cite{burgy-PRL.87.277202,dagotto-NJP2005,cen_c-PRL.98.127202}.
 In superlattices, however, non trivial phenomena may be observed and ordered hetero-valent ions do not prevent the emergence of a large magnetoresistance
 \cite{nakao-PRB.92.245104}. Particularly studied are (LaMnO$_3$)$_n${$\vert$}(SrMnO$_3$)$_m$ superlattices, where various magnetic and electronic ground states
 across the metal-insulator transition can be tuned
 \cite{adamo-PRB.79.045125,bhattacharya-PRL.100.257203,nanda-PRB.79.054428,smadici-PRL.99.196404,dong-PRB.78.201102,chen_h-JPCM2017}. This is achieved by varying
 their period ($n + m$) or component ratio ($n/m$) \cite{bhattacharya-PRL.100.257203,nanda-PRB.79.054428,may-nmat2009}, which is a way to modulate the tunnelling of
 $e_g$ electrons across the interface \cite{nanda-PRB.79.054428}; orbital \cite{nakao-PRB.98.245146} and charge \cite{pardo-APL2014} order were also reported.
 More exotic phenomena, including correlated topological states, may be expected when passing from the (001) orientation to the (111) orientation, because of the large
 polarity and a peculiar symmetry-driven epitaxial strain~\cite{myself-EPL2017,chakhalian-APLMat2020}. This is the reason why (111)-oriented superlattices remain under
 intense scrutiny, despite of the scarcity of suitable substrates and unfavourable thermodynamics \cite{chakhalian-APLMat2020,mastrikov-SS2008,mantz-SS2020}.

  In this context, the current article presents an \textit{ab-initio} study on the structural, electronic and magnetic properties of a (111)-oriented
 (LaMnO$_3$)$_{12}\vert$(SrMnO$_3$)$_{6}$ superlattice as illustrated in Fig.\ \ref{fig:proj-dos}(a), which is isostochiometric to the colossal
 magnetoresistive La$_{2/3}$Sr$_{1/3}$MnO$_{3}$. Our calculations will show that this superlattice has a half-metallic ferromagnetic (FM) ground state,
 whose character persists inside the innermost layers of the component regions. This behaviour is profoundly different from what reported for other
 orientations and is traced back to the cooperation of charge transfer across the interface, strain, structural distortions and electronic
 correlations. Magnetism will be shown to originate from a double-exchange interaction  between the Mn atoms and to be pinned inside the LaMnO$_3$
 region and not at the interface. Finally, the Mott-Hund character of the electronic correlations will also be analysed.

\section{Results and discussion}
\begin{figure}[t]
\centering
\includegraphics[trim= 1.0cm 0.0cm 2.0cm 0cm,width=\textwidth,height=0.8\textheight,keepaspectratio]{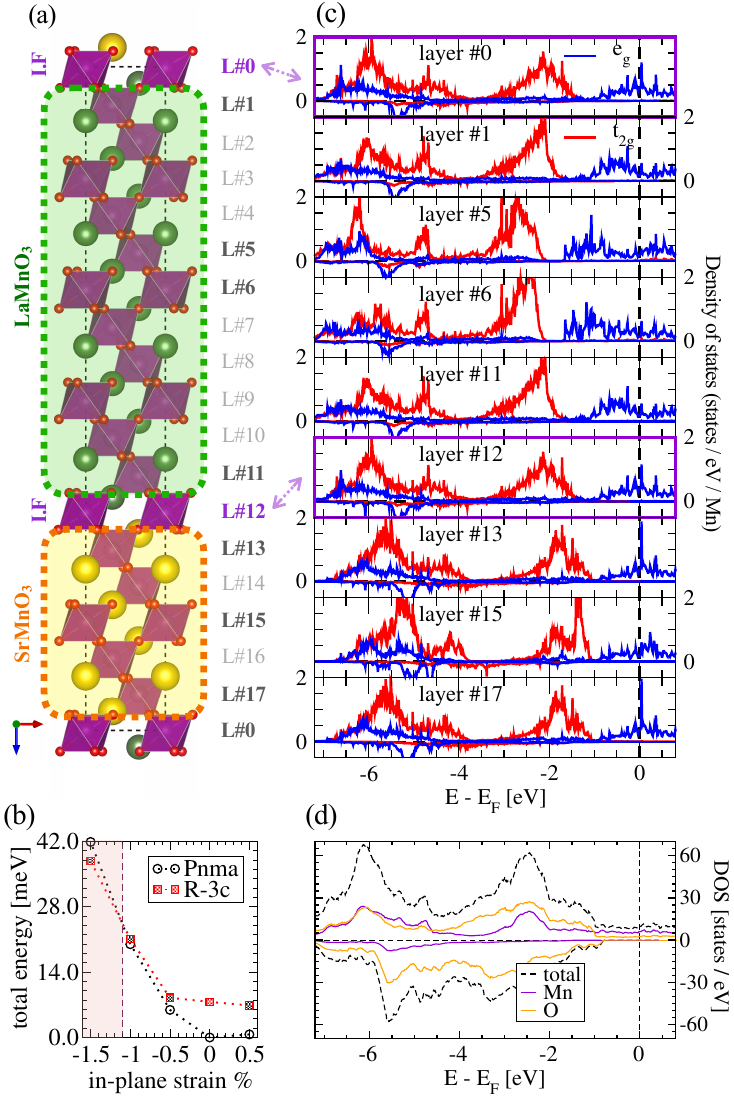}
	\caption{(a) Crystal structure of the superlattice. The interfacial layers
	are \#0 and \#12. (b) Curve total energy vs strain for FM solutions with
	$a^{-}a^{-}c^{+}$ (black) and $a^{-}a^{-}a^{-}$ (red); `zero' strain means
	that the pseudo-cubic lattice constant is \SI{3.860}{\angstrom}. (c) Projected
	DOS for Mn-$t_{2g}$ and Mn-$e_{g}$ states for selected layers; additional
	information, covering all Mn and O layers, is provided in the Supplemental
	Material. (d) Total, Mn-$d$ projected and O-$p$
	projected DOS for the FM solution.}
\label{fig:proj-dos}
\end{figure}

 {\bf{Ground-state structure and spectral properties.}}
  The (LaMnO$_3$)$_{12}\vert$(SrMnO$_3$)$_{6}$ superlattice is investigated via density functional theory (DFT) plus Hubbard U approach, labelled as
 sDFT+U. While tilting systems and angles are known for many perovskites in the bulk, their determination at surfaces and interfaces is not trivial.
 A previous \textit{ab-initio} study~\cite{myself-EPL2017} predicted that (111)-oriented manganite superlattices should adopt the $a^{-}a^{-}a^{-}$
 tilting system instead of the native $a^{-}a^{-}c^{+}$ tilting system, as expressed in Glazer notation \cite{glazer-AC:B1972}, used throughout this
 paper. The energy difference between these two structures depends on the lattice parameters. With a lattice constant of \SI{3.860}{\angstrom} (hereby
 denoted as equilibrium or 0 \% strain), the $a^{-}a^{-}c^{+}$ tilting system is favoured by \SI{7.6}{\milli\electronvolt} per formula unit (f.u., i.e.
 a AMnO$_3$ unit, where A may be La or Sr) with respect to the $a^{-}a^{-}a^{-}$ tilting system. A reasonably small compressive strain may change this
 structural order, as recently reported for nickelates \cite{kim_jr-ncomm2020}. As shown in Fig.\ \ref{fig:proj-dos}(b), we predict a transition to the
 $a^{-}a^{-}a^{-}$ tilting system for a compressive strain of $\sim 1.5$~\%, corresponding to epitaxial growth on SrMnO$_3$ (on LaAlO$_3$ the strain is
 $\sim 2$~\%). The energy difference between the two tilting systems remains below \SI{7}{\milli\electronvolt} (\SI{\sim 80}{\kelvin}) at all simulated
 strain values, whereas that between the ground state FM and anti-ferromagnetic (AFM) orders are larger than \SI{25}{\milli\electronvolt}
 (\SI{\sim 290}{\kelvin}), see Table \ref{tab:toten_tm}. Therefore, phase transitions and coexistence are more likely to occur in the structure than in
 the magnetic order.

 The two tilting systems do not lead to qualitatively different results in terms of distribution of charges and magnetic moments, but the
 $a^{-}a^{-}c^{+}$ requires a large supercell, which may hinder a thorough analysis of the layer-resolved properties. Therefore, we focus
 on the $a^{-}a^{-}a^{-}$ tilting system at the equilibrium lattice constant, for simplicity.

  The superlattice analysed has a half-metallic character, as inferred by the total density of electronic states (DOS) in Fig.\ \ref{fig:proj-dos}(d), with
 a band-gap in the minority-spin channel between Mn-$d$ and O-$p$ of \SI{\sim 2.0}{\electronvolt}. The site-projected DOS in Fig.\ \ref{fig:proj-dos}(c)
 reveals that the half-metallicity persists across all the layers. The curves show that the deeper the Mn inside the LaMnO$_3$ region, the lower the onset
 of the $e_{g}$ bands, because of the larger electronegativity of La with respect to Sr. The smooth variation of these onsets across the superlattice is a
 further signature of metallicity. Moreover, the $e_{g}$ states are well separated from the $t_{2g}$ states in the LaMnO$_3$ region (see layers 5 and 6),
 whereas they are much closer in the SrMnO$_3$ region. Their separation is small also at the interface, note the larger upshift of the $t_{2g}$ states with
 respect to the $e_{g}$ states in Fig.\ \ref{fig:proj-dos}(b).

 The features just outlined are unusual for manganite superlattices with this large thickness. For the (001) orientation, the superlattice becomes insulating
 if LaMnO$_3$ is thicker than 2 layers \cite{nanda-PRB.79.054428,bhattacharya-AnnuRevMR2014,pardo-APL2014}, whereas it is 12-layers thick in the present case. To understand
 the difference between the two orientations we should first understand what drives the formation of a half-metallic FM state. We recall that bulk LaMnO$_3$ (characterised by
 the $a^{-}a^{-}c^{+}$ tilting system) becomes half-metallic FM under a small compressive strain \cite{lee_jh-PRB.88.174426,rivero-PRB.93.094409}. As discussed above, SrMnO$_3$
 causes a small compressive strain on LaMnO$_3$, and this should be sufficient to induce the transition. Interestingly, this situation is even more favourable in the
 $a^{-}a^{-}a^{-}$ tilting system, which is predicted by our calculations to be a FM half-metal for all in-plane lattice parameters hereby considered.

  Once established that strain is the driving factor in determining the half-metallic FM state in the LaMnO$_3$ region, we need to understand why this state
 is more likely to survive in the (111)-oriented superlattice than in its (001)-oriented counterpart. Clarifying this issue requires a deeper analysis of the
 structural features and the magnetic properties, presented in the next sections.

\begin{figure}[ht]
\centering
\includegraphics[width=1.0\linewidth]{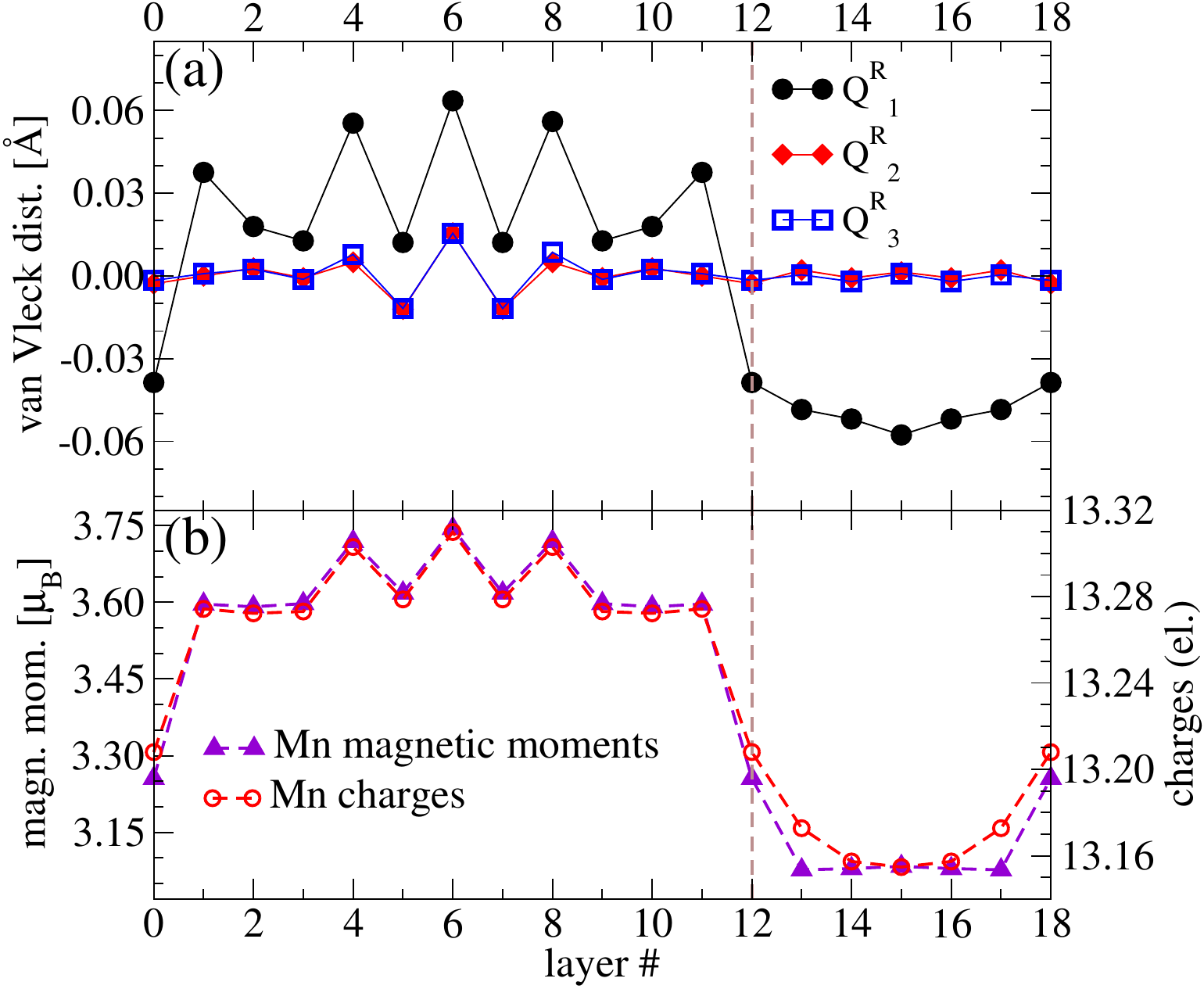}
	\caption{(a) Layer-resolved van Vleck distortions. (b) Mn-projected
	 Bader charges and magnetic moments. The dashed vertical line indicates
	 the interfacial layer. The charge transfer across the interface is
	 visible in having Bader charges larger (lower) than for the nominal
	 oxidation state, i.e. 13 (14) in the SrMnO3 (LaMnO3) region. A more
	 direct visualization of the structure with respect to the layers
	 numbering is provided in Fig.\ \ref{fig:proj-dos}.}
\label{fig:mn_mag}
\end{figure}

{\bf{Breathing distortions and spin-charge oscillations.}}
  As for most perovskites, the structural features are linked to the electronic and magnetic properties. In
 agreement with leading literature \cite{schmitt-PRB.101.214304,vanvleck-JChemPhys1939}, we introduce the
 Jahn-Teller distortions and breathing distortions in terms of the variations of the octahedral lengths $x,y,z$
 (with respect to their average values). Breathing distortions are defined as
 $Q^R_1 = (\Delta x + \Delta y + \Delta z)/\sqrt{3}$, whereas the Jahn-Teller distortions are
 $Q^R_2 = (\Delta x - \Delta y)/\sqrt{2}$ and $Q^R_3 = (- \Delta x - \Delta y + 2 \Delta z)/\sqrt{6}$.
 Breathing distortions are seldom found in manganites, which host orbital order and Jahn-Teller distortions
 (also defined in the Methods) instead. A recent study highlighted that Jahn-Teller modes arise from a steric
 effect that affects the electron-lattice coupling and are therefore dependent on the tilting system
 \cite{varignon-PRB.100.035119,varignon-PRR.1.033131}. In the bulk, the constraint imposed by the $R\bar{3}c$ phase
 should lead to a total quenching of these (pseudo) Jahn-Teller modes. The Jahn-Teller distortions are shown in Fig.\
 \ref{fig:mn_mag}(a) and are quenched in agreement with the aforementioned literature. The quenching is not full
 because the relaxation of the superlattice modifies the pristine $a^{-}a^{-}a^{-}$ tilting pattern.

  The quenching of the Jahn-Teller modes is accompanied by the presence of the breathing modes. The latter are lessened by a factor 4 in the structure without tilts (data
 not shown). A similar relation between octahedral tilts and breathing distortions was recently found in rare-earth nickelates, where it leads to a structurally triggered
 metal-insulator transition \cite{mercy-ncomm2017}. In addition, LaMnO$_3$ is mentioned as a case where a close competition between charge and orbital order is driven by a
 similar mechanism (in line with Refs.\ \onlinecite{schmitt-PRB.101.214304} and \cite{varignon-ncomm2019}). In the superlattice under investigation, this mechanism has to
 compete with the high stability of the FM half-metallic phase, associated to the strained structure, and with the uniform shift of the band-edge, induced by the charge
 transfer -- see again Fig.\ \ref{fig:proj-dos}(b). Therefore, it becomes unfavourable to induce a transition to an AFM insulator with orbital order. The site-projected
 charge and magnetic moment distributions, as computed {\em \`a la} Bader \cite{tang_w-JPCM2009,sanville-JCompChem2007,henkelman-CMSci2006,yu_m-JChemPhys2011,kerrigan_a-github},
 is presented in Fig.\ \ref{fig:mn_mag}(b) and shows a hint of charge order, leading to oscillations in the LaMnO$_3$ region and a smooth behaviour in the SrMnO$_3$ region.

\begin{figure}[t]
\centering
\includegraphics[width=1\linewidth]{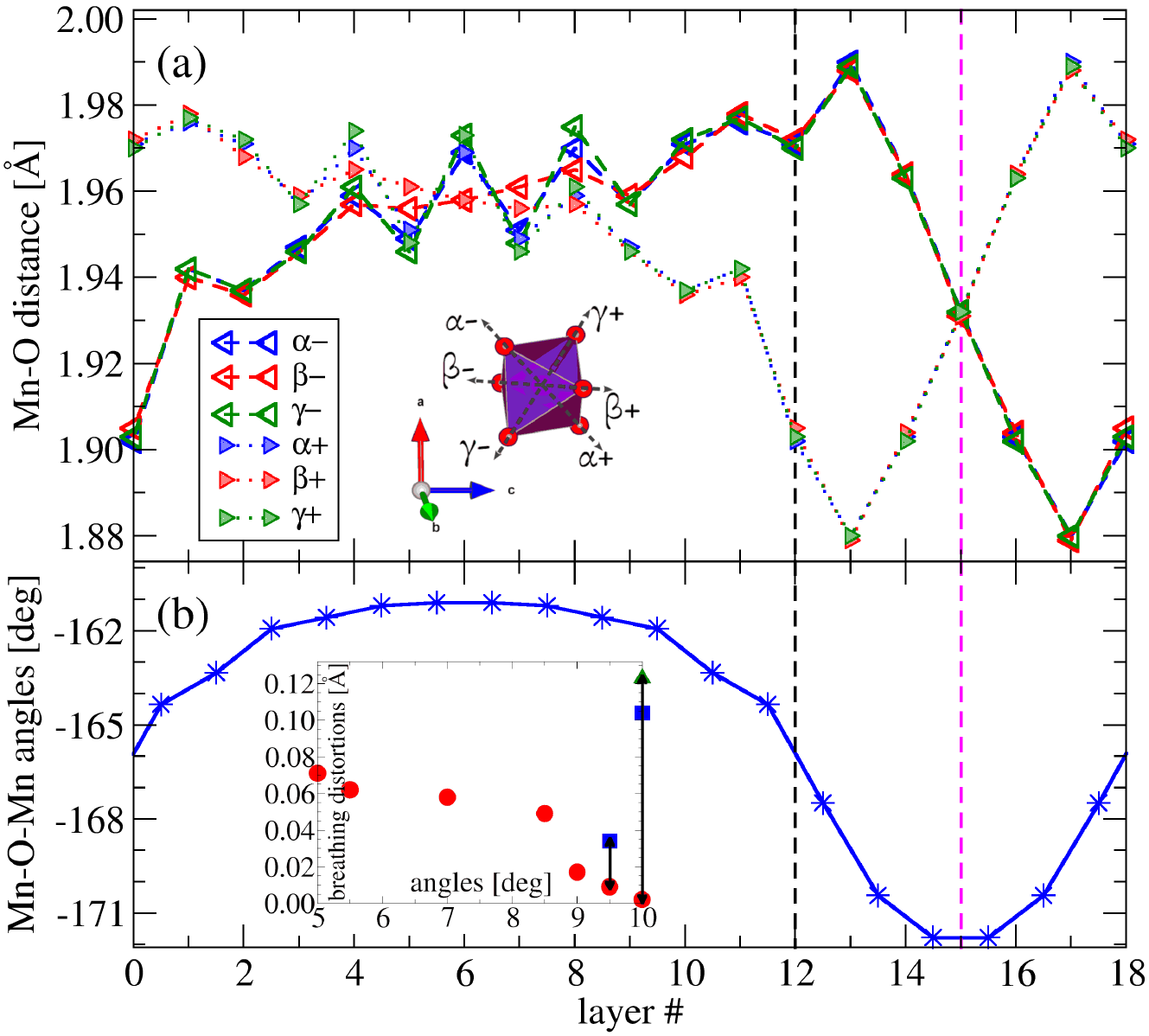}
	\caption{(a) Mn-O distance for the 3 inequivalent nearest neighbours,
	as illustrated in the inset; $+$ and $-$ denote increasing and decreasing
	$c$, respectively. (b) opposite of the Mn-O-Mn angles ($\widehat{MnOMn}$)
	averaged over the three directions; inset: correlation between the angles
	$(\pi - \widehat{MnOMn})/2$ and the breathing distortions. In the main
	panels, the vertical dashed black line denotes the layer at the interface,
	while the dashed magenta line denotes the center of the SrMnO$_3$ region.
	Note that layers \#0 and \#18 are also at the interface.In the inset of
	panel (b), the vertical lines denote the range of breathing distortions
	for one corresponding value of the angle.}
\label{fig:bondlang}
\end{figure}

  The smooth variation of the Mn-O-Mn angles across the superlattice (which take approximately the same values for all layers in the central
 LaMnO$_3$ region) mirrors the uniform shift of the band-edge, compare Figs.\ \ref{fig:bondlang}(b) and \ref{fig:proj-dos}(b), whereas the
 Mn-O distances appear with oscillations in the inner LaMnO$_3$ region, see Fig.\ \ref{fig:bondlang}(a). Where the Mn-O-Mn bond are closer to
 a flat angle (the SrMnO$_3$ region) the structure presents larger splitting in the Mn-O distances between increasing $\hat{z}$ and decreasing
 $\hat{z}$, and where the Mn-O-Mn bonds are more bent (the LaMnO$_3$ region) the structure presents oscillating Mn-O distances and breathing
 modes, see Figs.\ \ref{fig:mn_mag} and \ref{fig:bondlang}. In fact, the connection between breathing distortions and angles is even more
 apparent in the inset of Fig.\ \ref{fig:bondlang}(b), where flat angles correspond to single values for the MnO$_6$ volumes (breathing
 distortion) and bent angles -- in the LaMnO$_3$ region -- correspond to large variance of the MnO$_6$ volumes distribution, in line with
 the above mentioned literature \cite{mercy-ncomm2017}. The large Mn-O distance splitting, occurring in the SrMnO$_3$ region, reveals a
 typical tendency for SrMnO$_3$ to ferroelasticity \cite{marthinsen-MRSComm2016}, avoided by the symmetry with respect to the interface.

\begin{figure*}
\centering
\includegraphics[trim= 1.0cm 0.0cm 0cm 0cm, width=1.0\linewidth]{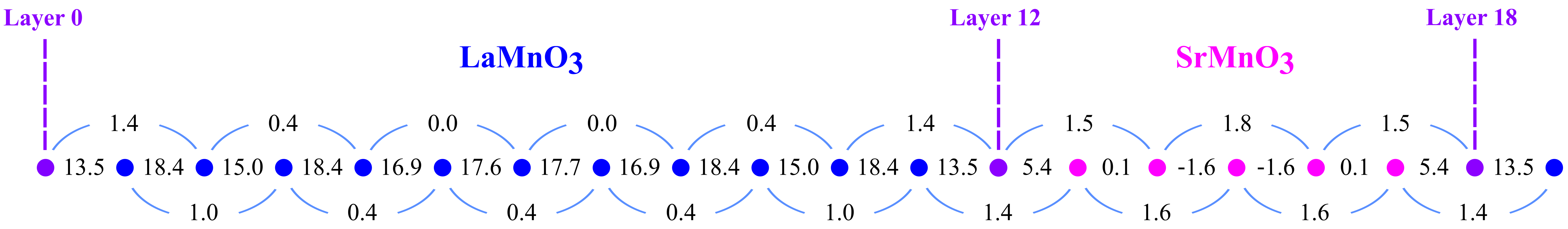}\\
	\caption{Magnetic coupling constants as computed and shown along the (001) -- or any of
	the equivalent (010) and (100) -- crystallographic directions of the superlattice. The
	Mn are shown as circles, and the values of the coupling constants are reported in meV.
	Straight segments denote first nearest neighbours, whereas semicircles denote fourth-nearest
	neighbours, namely Mn two layers away but along the same Mn-O-Mn direction (whereas they
	vanish for all other Mn couples).}
\label{fig:jij}
\end{figure*}

{\bf{Exchange couplings and ferromagnetic order.}}
  As we argued that the magnetism is a consequence of structure and strain, we do not expect it to be interfacially driven. This is clearly visible in Fig.\
 \ref{fig:mn_mag}(b), where the largest magnetization is found in the innermost layers of the LaMnO$_3$ region. For a better insight into the magnetic properties,
 we analyse the inter-atomic exchange coupling, computed for a lattice constant of \SI{3.892}{\angstrom} (which corresponds to a tensile strain below 1\%). The
 largest contributions in the Mn sublattice are those connecting a Mn atom to its first nearest neighbours or fourth-nearest neighbours. The latter correspond
 to the second-nearest neighbours along Mn-O-Mn-$\cdots$ lines, consistently with the double-exchange mechanism~\cite{nanda-PRB.79.054428,bhattacharya-AnnuRevMR2014}.
 The relevant exchange couplings across the superlattice are illustrated in Fig.\ \ref{fig:jij}. In the LaMnO$_3$ region, the magnetic order is driven by the FM
 nearest neighbour coupling, which in the innermost layer takes the value of 17.7 meV and sharply decreases at the interface, exhibiting oscillations in phase with
 the magnetic moments. Interestingly, the maximum value is not reached at the innermost layer, but at intermediate layers, and amounts to \SI{18.4}{\milli\electronvolt},
 which is 30\% smaller than in the isostochiometric La$_{2/3}$Sr$_{1/3}$MnO$_{3}$~\cite{furrer-PRB.95.104414}. This behaviour reflects a competition between the
 trend of the magnetization -- see Fig.\ \ref{fig:mn_mag}(b) -- and the potential induced by the charge transfer across the interface -- see Fig.\ \ref{fig:proj-dos}(b).
 Such relatively strong ferromagnetism is even more surprising if compared to the behaviour of (001)-oriented supercells, whose nearest neighbour exchange becomes
 bulk-like AFM for LaMnO$_3$ regions thicker than 2 unit cells~\cite{nanda-PRB.79.054428,bhattacharya-AnnuRevMR2014}. A smaller contribution to the magnetic order
 is given by the fourth-nearest neighbour exchange, whose values are noticeable at the interface (1.36 meV), but are totally quenched in the innermost layers of the
 LaMnO$_3$ region.

  The situation is more complicated in the SrMnO$_3$ region. In the innermost layers, the nearest neighbour exchange is AFM, as in the bulk
 \cite{ricca-PRB.99.094102}. However, The strength of the coupling is much weaker than in the bulk, i.e. \SI{-1.6}{\milli\electronvolt} versus
 \SI{-7.5}{\milli\electronvolt} \cite{zhu-PRB.101.064401}, due to the combined effect of charge transfer and epitaxial strain (about 1$\%$).
 Strain alone was shown to induce an AFM-FM transition at about 3\% in bulk cubic SrMnO$_3$~\cite{zhu-PRB.101.064401} -- while here the strain
 is virtually null on SrMnO$_3$. Interestingly, the FM order inside the SrMnO$_3$ region is stabilised by the fourth-nearest neighbour coupling,
 which becomes even larger (\SI{1.8}{\milli\electronvolt}) than the nearest neighbour one. This frustration due to competing FM and AFM
 interactions is likely to lead to a more complex magnetic structure, probably accompanied by non-collinearity. Exploring the magnetic phase
 diagram may be an interesting project, but outside the scope of the present work. We prefer, instead, to focus on the origin of the oscillations
 of charge, magnetic moments, exchange couplings and breathing distortions.

\begin{figure*}
\centering
\includegraphics[trim= 0cm 0.0cm 0cm 0cm, width=1.0\linewidth]{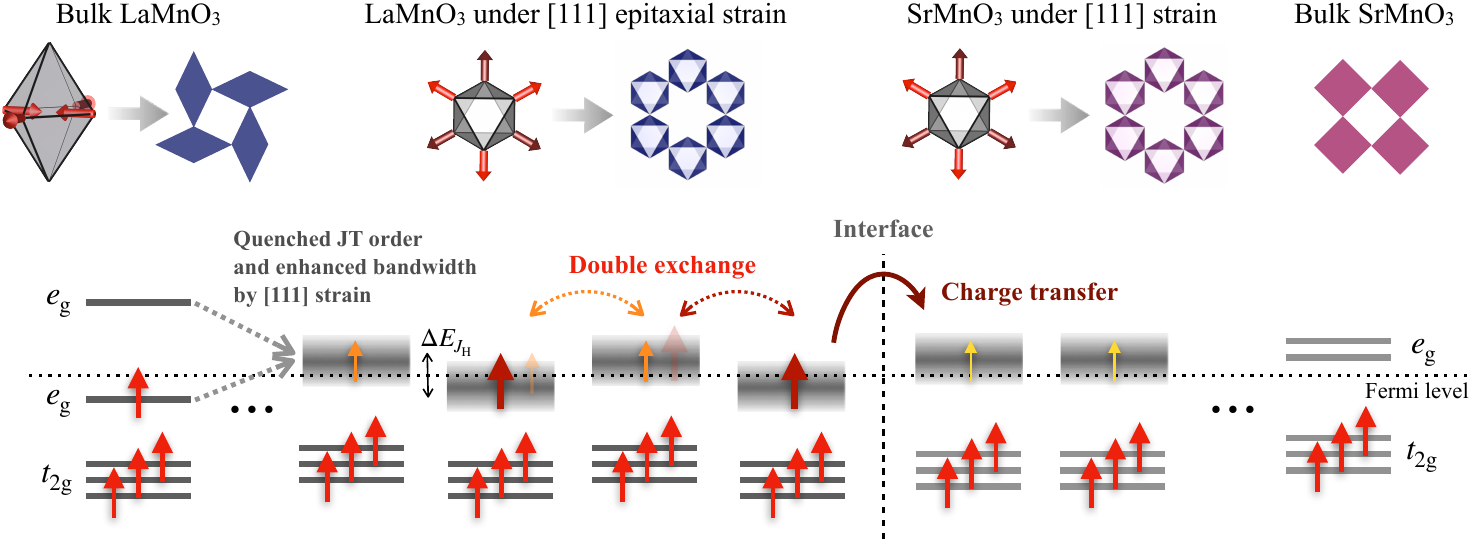}
	\caption{Sketch of the physical mechanisms driving the interface behaviour in the superlattice. Bulk LaMnO$_3$
	(left side) is an AFM Mott insulator with Jahn-Teller distortion and orbital order, in a $d^4$ ionic configuration.
	Bulk SrMnO$_3$ (right side) is an AFM band insulator in a $d^3$ configuration. Across the interface, charge
	transfer and epitaxial strain lead to partially filled bands and breathing distortions, while the Jahn-Teller
	order is quenched. Metallicity favours the FM order driven by the double-exchange mechanism, and in turn accompanied
	by spin and charge oscillations.}
\label{fig:sketch}
\end{figure*}

  We can summarise what leads to the half-metallic FM state with the help of Fig.\ \ref{fig:sketch}. In the bulk, LaMnO$_3$ and SrMnO$_3$ behave as an AFM Mott
 insulator with Mn$^{3+}$ ions and an AFM band insulator with Mn$^{+4}$ ions, respectively. In the superlattice, the local strain in the LaMnO$_3$ region induces
 the delocalisation of the Mn-3d states, which in turn suppresses the AFM super-exchange and favours the FM double-exchange
 \cite{bhattacharya-AnnuRevMR2014,lee_jh-PRB.88.174426}. This effect is further enhanced by the charge transfer across the interface, which penalises the ionic
 picture and promotes the hopping between Mn sites. In (001)-oriented superlattices, the local strain is imposed in-plane -- hence along two crystallographic
 directions -- and allows different relaxations in different regions: SrMnO$_3$ recovers its G-type AFM order, blocking the tunnelling of $e_{g}$ electrons from
 LaMnO$_3$ and imposing a strong penalty on the double-exchange mechanism. For the (111) orientation, the strain acts on the same footing for all octahedral axes,
 and therefore the aforementioned phase separation is forbidden, the $e_{g}$ tunnelling survives and the FM coupling prevails. In summary, geometrical degrees of
 freedom affect the electronic ones, governing the magnetic and metallic properties of the superlattice. Further information can be inferred by the analysis of the
 bond angles, shown in Fig.\ \ref{fig:bondlang}. In the LaMnO$_3$ region, the Mn-O-Mn angles vary from 160{\textdegree} to 165{\textdegree}. These values are higher
 than the bulk LaMnO$_3$ \cite{jilili-SciRep2015} value of 155{\textdegree} and close to the La$_{2/3}$Sr$_{1/3}$MnO$_{3}$ value \cite{colizzi-PRB.78.235122} of
 165{\textdegree}. The analysis of the Mn-O bond lengths, see Fig.\ \ref{fig:bondlang}, is in line with the presence of breathing modes in the LaMnO$_3$, but also
 with a tendency to ferroelasticity in SrMnO$_3$, avoided by the equivalence of the interfaces. Such behaviour is typical of SrMnO$_3$ \cite{marthinsen-MRSComm2016}.
 Charge doping and metallicity would anyway prevent the transition to a ferroelectric phase.

\begin{figure}
\centering
\includegraphics[width=1.0\linewidth]{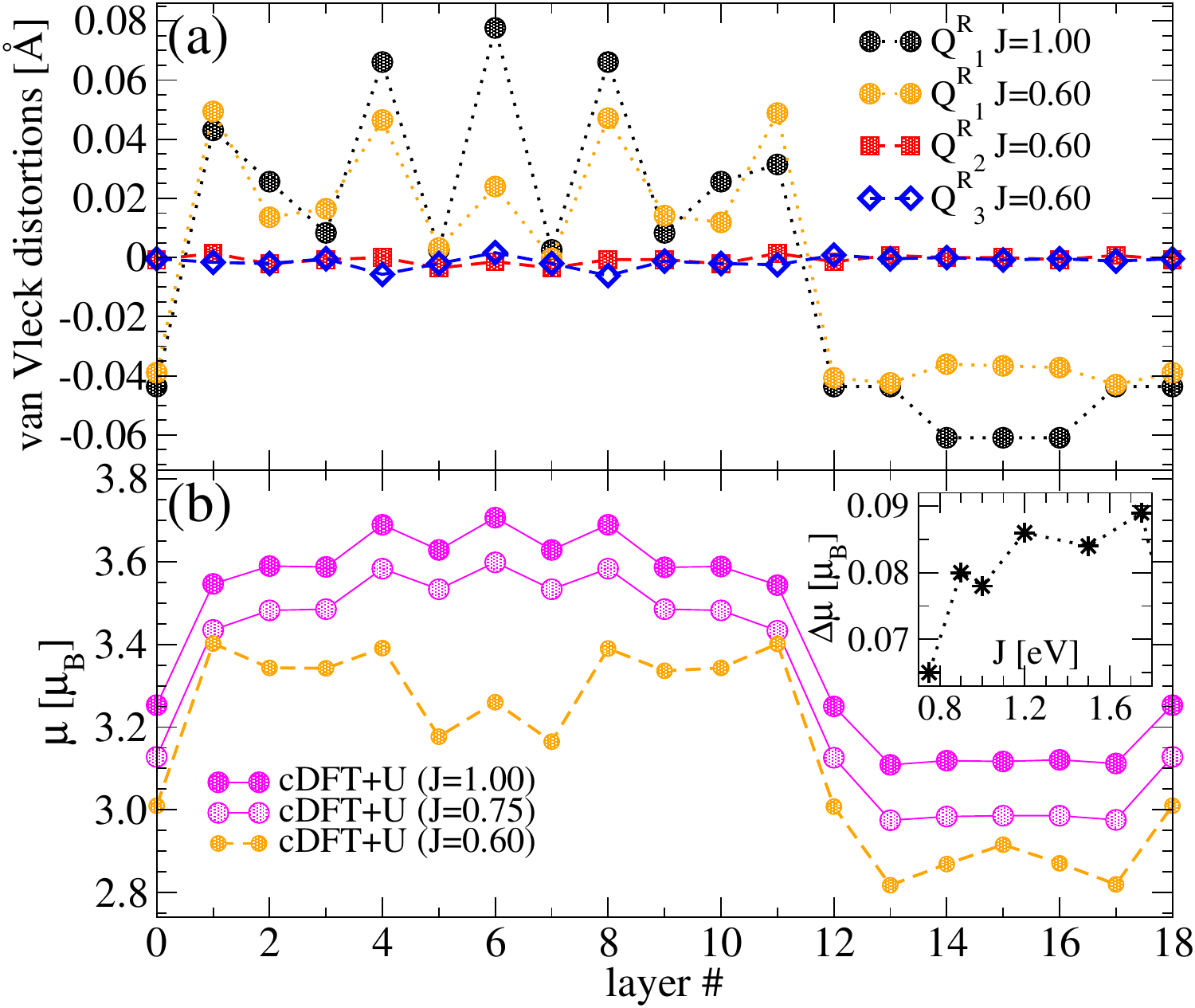}
	\caption{(a) Layer-resolved van Vleck distortions and (b) Mn-projected Bader magnetic moments
	 for different values of Hund's exchange $J$ (in eV). See the Methods section for details on
	 cDFT+U$\vert$J and sDFT+U$\vert$J. Inset: difference $\Delta\mu$ between the magnetic moments
	 at layers \#5 and \#6, which provides a measure of the amplitude of the moment oscillations.}
\label{fig:hunds_dependence}
\end{figure}

{\bf{Role of Hund's coupling.}}
  The suppression of Jahn-Teller order in favour of breathing distortions was predicted a decade ago in nickelates as a consequence
 of Hund's coupling and was pointed out to be persistent well into the metallic side of the Mott transition \cite{mazin-PRL.98.176406}.
 Later studies \cite{mercy-ncomm2017} pointed to structural distortions as the driving mechanism. In the present case, we observe an
 intermediate situation. On one hand, the structure has a primary importance, forbidding certain distortions, such as Jahn-Teller,
 and preserving metallicity. On the other hand, the breathing distortions are not large enough to induce a metal-insulator transition,
 but we still observe signs of charge order. The dependence of our results on the strength of Hund's exchange $J$ is investigated via
 charge density functional plus $U$ scheme (denoted as cDFT+U$\vert$J hereafter) \cite{keshavarz-PRB.97.184404,jang_sw-PRB.98.125126}
 for a lattice constant of \SI{3.892}{\angstrom}. It suggests to what extent Hund's coupling affects the properties of our system. The
 oscillations of the magnetic moments across the superlattice depend crucially on the value of Hund's coupling, see Fig.\
 \ref{fig:hunds_dependence}(b). When $J$ becomes as small as \SI{0.6}{\electronvolt}, the magnetism is no longer pinned at the innermost
 layers of the LaMnO$_3$ region. Instead, it becomes pinned at the interface, similarly to what happens in (001)-oriented superlattices.
 The change in the trend of the magnetic moments across the superlattice is accompanied by an analogous change of the breathing distortions,
 emphasising that the former drives the latter, to a large extent. Furthermore, electron/hole doping may lead to the disappearance of the
 breathing distortions as well as the magnetic moments oscillations (data not shown). Such a drastic change is surprising, considering the
 shape of the density of states in the corresponding doping range ($\pm 0.1$ eV), and suggests that strong electronic correlations play an
 important role. Overall, the metallic character with spin and charge oscillations, the presence of strong correlation effects, and the key
 role of the Hund's exchange $J$ suggest that our superlattice behaves as a Hund's metal \cite{georges-AnnuRevCMP2013,stadler-AnnPhys2019365}.
 Using our parameters $U$ and $J$, as well as the effective bandwidth extracted from the DOS, we can obtain some information from existing
 phase diagrams of the Hubbard model \cite{fanfarillo-PRB.92.075136,isidori-PRL.122.186401,merkel_me-PRB.104.165135}. The most accurate
 comparison is offered by Ref.\ \onlinecite{merkel_me-PRB.104.165135}, where Merkel \textit{et al.} investigated a 5 orbital system with
 a $d^{4}$ occupation, including the level splittings associated to the presence of breathing modes (representative of CaFeO3). Using their
 phase diagram, we can confirm the regime of (homogeneous) Hund's metallicity, with an estimated quasiparticle weight between 0.4 and 0.6.
 By increasing $J$, we expect to get closer to a valence skipping metal phase, which finds correspondence in an increasing amplitude of
 charge and spin oscillations (see inset of Fig.\ \ref{fig:hunds_dependence}(b)). A further increase of $J$ would lead to a charge
 disproportionated insulator, which in Ref.\ \onlinecite{merkel_me-PRB.104.165135} is predicted to happen for values larger than 2.3 eV.
 Despite signs of ``Hundness'' have been found in a variety of systems during the last decade \cite{georges-AnnuRevCMP2013,stadler-AnnPhys2019365},
 they had never been reported for a manganite superlattice. For a more quantitative analysis allowing for more precise conclusions, one would
 need to perform calculations beyond DFT, e.g. in combination with the dynamical mean-field theory (DMFT), including the structural response.
 Considering the size of the system as well as the very high number of degrees of freedom, such an analysis would be beyond the scope of the
 present study.

{\bf{Outlook.}}
  In summary, we have investigated the electronic and magnetic properties of the (111)-oriented
 (LaMnO$_3$)$_{12}\vert$(SrMnO$_3$)$_{6}$ superlattice using DFT+U$\vert$J. A half-metallic FM
 state is supported by the cooperation of charge, spin, orbital and lattice degrees of freedom,
 and is favoured with respect to an AFM state. The half-metallic FM character is found to persist
 across the entire superlattice, while the innermost layers of its (001)-oriented counterpart
 become AFM insulators for a thickness larger than 3 layers. The atomic volumes, charges and
 magnetic moments are correlated across the superlattice, in particular in the LaMnO$_3$ region,
 where adjacent sites display charge, spin, and volume oscillations. Breathing distortions arise
 and may be accompanied by the quenching of Jahn-Teller distortions in the presence of
 $a^{-}a^{-}a^{-}$ tilting of MnO$_6$ octahedra while may coexist with the Jahn-Teller distortions
 in other tilting systems such as $a^{-}a^{-}c^{+}$. Overall, the present results suggest that the
 [111] epitaxial strain associated to the superlattice formation is a viable pathway to engineer
 a system analogous to La$_{2/3}$Sr$_{1/3}$MnO$_{3}$ without introducing doping-induced disorder.
 This is not only an advantage for avoiding alloy-related problems, as e.g. cation disorder, but
 does also widen the potential of interfacial engineering in oxide heterostructures. Finally, we
 also speculate that the fascinating physics exhibited by this superlattice may arise from Hundness,
 alongside with structural aspects, similarly to what happens in nickelates. Therefore, the system
 addressed here can provide further theoretical ground for the development of heterostructures
 hosting exotic magnetic phases and topological states.

\section{Methods}

{\bf{Framework and parameters.}}
 The superlattice is fully relaxed with an optimised lattice constant of \SI{3.860}{\angstrom} (denoted as equilibrium or 0 \% strain).
 Calculations are performed in density functional theory (DFT) using the projector-augmented wave method as implemented in the Vienna
 {\itshape{Ab-initio}} Simulation Package (VASP) \cite{kresse-PRB.54.11169,kresse-PRB.59.1758}. The exchange-correlation functional is
 treated in the generalised gradient approximation (GGA) by Perdew–Burke–Ernzerhof \cite{perdew-PRL.77.3865,perdew-PRLerratum.78.1396}. 
 To improve the description of the Mn-3d states \cite{mellan-PRB.92.085151}, we make use of on-site corrections for static correlation
 effects in the rotational invariant DFT+U approach by Liechtenstein \textit{et al.} \cite{liechtenstein-PRB.52.r5467}, denoted as
 sDFT+U$\vert$J. Further calculations to analyse Hund's coupling are performed by following the approach discussed in Ref.\
 \onlinecite{park_hw-PRB.92.035146}, which we label as cDFT+U$\vert$J. The Coulomb interaction parameters are chosen as $U =$
 \SI{3.8}{\electronvolt} and $J =$ \SI{1.0}{\electronvolt}, in accordance with works on similar systems
 \cite{nanda-PRB.81.224408,nanda-PRL.101.127201}. A deeper analysis of the magnetic properties is then performed via the full-potential
 linear muffin-tin orbital (FP-LMTO) method as implemented in the RSPt code~\cite{RSPt_Springer2010,rspt_website,granas-CMSci2012}.
 We used also the SCAN parameter-free functional \cite{sun_jw-PRL.115.036402}: results (shown in the Supplemental Material) confirm the core
 results obtained with DFT+U$\vert$J.

  The inter-atomic exchange interactions $J_{ij}$ are calculated by mapping the magnetic excitations onto an effective Heisenberg Hamiltonian
 $\hat H = - \sum_{i \neq j} J_{ij} \cdot ( \vec{e}_i \cdot \vec{e}_j )$, where $i$, $j$ are atomic sites and $\vec{e}_i$, $\vec{e}_j$ are
 unit vectors along the local magnetization direction. This calculation is performed via the magnetic force theorem, using the implementation
 of Ref.\ \onlinecite{kvashnin-PRB.91.125133}, which was also successfully applied to CaMnO3 \cite{keshavarz-PRB.95.115120}.
 Due to the better accuracy of all-electron methods~\cite{lejaeghere-Science2016},
 these calculations also serve to confirm the validity of VASP results.

{\bf{General considerations on the modelled system.}}
  The supercells used for the calculations consist of two types: the $R\bar{3}c$ cell was used for the results presented throughout this work, whereas
 an orthorhombic cell was used for calculations to compare total energies and results on the charge and spin distributions. The two interfaces have the
 same stacking sequence (i.e. LaO$_3\vert$Mn$\vert$SrO$_3$ or SrO$_3\vert$Mn$\vert$LaO$_3$), thus they are equivalent; furthermore, the superlattice
 possess inversion symmetry with respect to the Mn atom at the interface and therefore there is no built-in electric field generated. Further details
 on the cells and the sampling of the Brillouin zones are given in the Supplemental Material.

  The analysis on the on-site charges and magnetic moments is carried on by in agreement with state-of-the-art methods
 \cite{tang_w-JPCM2009,sanville-JCompChem2007,henkelman-CMSci2006,yu_m-JChemPhys2011,kerrigan_a-github}. The analysis
 of the electronic properties is performed with the aid of the post-processing code VASPKIT \cite{VASPKIT-1908.08269}.
 Finally, images of structures and charge/spin distributions are produced with VESTA JP-Minerals \cite{VESTA-JACr2011}.
 Further details on the calculations are given in the Supplemental Material.

\section{Code availability}
  The calculations for this work have been performed with VASP and RSPt. The former is a licence product from the University of Vienna;
 the licence can be obtained upon submitting an application through the vasp portal (http://www.vasp.at). The latter is a free software
 distributed under GPL license after registration to a mailing list. More information about RSPt can be found at http://fplmto-rspt.org/
 \cite{rspt_website}.

\section{Data availability}
  The datasets generated during and/or analysed during the current study are available from the first author on reasonable request.

\section{Acknowledgments}
  We acknowledge the support of computational resources, including technical assistance, from the Swedish National Infrastructure for Computing
 (SNIC) at the National Supercomputer Centre (NSC) of Link\"oping University (Sweden) and at the High Performance Computing Centre North (HPC2N),
 partially funded by the Swedish Research Council through grant agreement no. 2018-05973, as well as from the National Supercomputing Center of
 Korea (Grant No.\ KSC-2020-CRE-0156). F.~C.\ acknowledges financial support from the National Research Foundation (NRF) funded by the Ministry of
 Science of Korea (Grant No. 2017R1D1A1B03033465). NRF support is also acknowledged by F.~C.\ and H.-S.~K.\ for the Basic Science Research Program,
 Grant No.\ 2020R1C1C1005900, and by I.~D.~M.\ for the Mid-Career Grant No.\ 2020R1A2C101217411. I.~D.~M.\ acknowledges also financial support from
 the European Research Council (ERC), Synergy Grant FASTCORR, Project No.\ 854843. The work of I.~D.~M.\ is supported by the appointment to the JRG
 program at the APCTP through the Science and Technology Promotion Fund and Lottery Fund of the Korean Government, as well as by the Korean Local
 Governments, Gyeongsangbuk-do Province and Pohang City. The authors, in particular F.~C., are also grateful to Hassan A.\ Tahini and Adam M.\
 Kerrigan for substantial scientific discussions and technical support.
 
 \section{Competing Interest}
 The authors declare that they have no competing interests.
  
 \section{Author contribution}
  F.~C.\ planned the project and performed all the VASP calculations. I.~D.~M.\ performed all the RSPt calculations. F.~C.\ and I.~D.~M.\
 wrote the initial manuscript. All authors contributed to analysing the data, revising the manuscript and drawing the conclusions.
 
 \section{Correspondence}
  Correspondence should be addressed to Heung-Sik Kim and Igor Di Marco. Emails: {\it{heungsikim@kangwon.ac.kr}} and {\it{igor.dimarco@apctp.org}}.

 \section{Figure Legends and Tables}

\begin{table}[b]
  \centering
  \caption{Energy difference per formula unit relative to the
	ground state structure -- i.e. the FM phase with the
	$a^{-}a^{-}c^{+}$ tilting system -- of various magnetic
	orders and tilting systems. The energies are computed
	for the same in-plane lattice parameters, corresponding
	to 0\% strain.}
        \begin{tabular}{rcc}
                \hline\hline
                    & $a^{-}a^{-}c^{+}$ & $a^{-}a^{-}a^{-}$ \\
                \cline{2-3}
                FM  &       G. S.       &      $  7.6$      \\
         A-type AFM &      $ 26.5$      &      $ 37.5$      \\
         C-type AFM &      $ 43.7$      &      $ 53.8$      \\
         G-type AFM &      $ 71.8$      &      $ 95.6$      \\
    \end{tabular}
\label{tab:toten_tm}
\end{table}

\end{document}